\documentclass[11pt]{article}

\usepackage{multirow}
\usepackage{xcolor}

\usepackage[english]{babel}

\usepackage{blindtext}
\usepackage{tikz}
\usetikzlibrary{backgrounds}

\usepackage[linesnumbered,boxed]{algorithm2e}
\usepackage{mathrsfs}
\usepackage{latexsym,lineno}
\usepackage{epsfig}
\usepackage{amsmath}\usepackage{fleqn}\usepackage{verbatim}\usepackage{epsf}
\usepackage{amsthm}\usepackage{graphicx, float}\usepackage{graphicx}
\usepackage{amsfonts}\usepackage{amssymb}\usepackage{graphpap}
\usepackage{epic}\usepackage{curves}

\numberwithin{equation}{section}

\title{\bf Algebro-geometric solution of the coupled
Burgers equation
}
\author{  Hongfei Pan   ,\,Tiecheng Xia\footnote{{\it Corresponding
author.} E-mail address: xiatc@shu.edu.cn(T. Xia).}  \\
{\small \it Department of Mathematics, Shanghai University, Shanghai
200444, China}
\\
}
\date{}
\begin{document}
\maketitle
\begin{abstract}
We derive theta function representation of algebro-geometric
solution of a Coupled Burgers equation which the second nonlinear
evolution equation in a hierarchy. We also derive the
algebro-geometric characters of the meromorphic function $\phi$ and
the Baker-Akhiezer vector $\Psi$.\\
\vspace{0.5mm} \textbf{keywords}: algebro-geometric solution;
algebraic curves; Riemann theta functions

\bigskip


\end{abstract}

\section{Introduction}

According to inverse spectral theory and algebro-geometric methods,
we can construct the explicit theta function representations of
quasi-periodic solutions of integrable nonlinear evolution
equations(including, soliton solutions as special limiting cases or
quasi-periodic behavior of nonlinear phenomenon)
\cite{Novikov-1980,Belokolos-1994,Gesztesy-2003,Gesztesy-2008}, and
this approach developed by pioneers such as Novikov, Dubrovin,
Mckean, Lax, Cao
\cite{Novikov-1974,Dubrovin-1975,Lax-1975,Mckean-1975,Cao-2003}. How
to obtain explicit solutions of the soliton equations, to reveal
inherent structure of soliton equations and to describe the
characteristic for the integrability of soliton equations, there are
also many works had been done
\cite{Ablowitz-1981,Hirota-2004,Cao-1990}.

In this paper, we will focus on a derivation of a coupled system and
search for the algebro-geometric solution of the following on the
basis of approaches in \cite{Gesztesy-2003,Hou-2013,Xue-2012}:
$$
\begin{array}{ll}
u_{t}  =  v_{xx}-2(uv)_{x},\\
v_{t}  =  u_{xx}+2uu_{x}-6vv_{x},
\end{array}\eqno(1.1)
$$
obviously, when $ u = v $, this system is the Burgers equation

$$u_{t} + 4uu_{x}- u_{xx}  =  0 ,$$ so we call the above system(1.1)
as the coupled Burgers equation.

The original motivation of this paper was to construct
algebro-geometric solution of the coupled Burgers equation based on
its obtained Lax pairs. The paper is organized as follows. In Sect.
2, we describe our zero-curvature formalism and derive the coupled
Burgers equation. In Sect. 3, we establish a direct relation between
the elliptic variables and the potentials. Our principal Sect. 4 is
devoted to a detailed derivation of theta function formulas of all
algebro-geometric quantities involved.

\section{The hierarchy and Lax pairs of the coupled Burgers equation }

In this section, we introduce the Lenard gradient sequence $
\{G_{j}\}_{ j = -1,0,1\dots }$  to derive the hierarchy associated
with equations (1.1) by the recursion relation
$$KG_{j-1}= JG_{j},\,\,\ j= 0,1,2,\ldots .\,\,\ G_{j}\arrowvert_{(u,v)=0}=0, \,\,\ G_{-1}= (1,1)^{T}, \eqno(2.1)$$
where $G_{j}=(G_{j}^{(1)}, G_{j}^{(2)})$ and two skew-symmetric
operators ($\partial = \partial/\partial x$)
$$
\begin{array}{ccc}
{K}= \left(
\begin{array}{cc} 0 & \partial^{2}- 2\partial u \\
 - \partial^{2}-2u \partial & -2 \partial v - 2v
\partial\
\end{array}\right)
\end{array},
\begin{array}{ccc}
{J}= \left(
\begin{array}{cc} 2\partial & 0 \\
 0 & -2 \partial\
\end{array}\right)
\end{array}. \eqno(2.2)
$$
A direct calculation gives from the recursion
relation(2.1)that$$G_{0}= (-u, v)^{T},\,\,\ G_{1}=(
\frac{1}{2}v_{x}- uv, -\frac{1}{2}u_{x}- \frac{1}{2}u^{2}+
\frac{3}{2}v^{2})^{T}. \eqno(2.3)$$

Consider the spectral problem
$$\psi_{x}= U \psi , \,\,\ \begin{array}{ccc}
{U}= \left(
\begin{array}{cc} -\lambda + v & u + v \\
 u - v & \lambda - v \
\end{array}\right)
\end{array} \eqno(2.4)$$
and the auxiliary problem:
$$\psi_{t_{m}}= V^{(m)} \psi , \,\,\ \begin{array}{ccc}
{V^{(m)}}= \left(
\begin{array}{cc} V_{11}^{(m)} & V_{12}^{(m)} \\
 V_{21}^{(m)} & -V_{11}^{(m)} \
\end{array}\right)
\end{array}, \eqno(2.5)$$
where
$$ V_{11}^{(m)}= -\sum_{j=0}^m 2(-\lambda +
v)G_{j-1}^{(2)}\lambda^{m-j} - \sum_{j=0}^m G_{j-1,x}^{(1)}
\lambda^{m-j},$$
$$ V_{12}^{(m)}= \sum_{j=0}^m [(G_{j-1,x}^{(2)}-
G_{j-1,x}^{(1)})-2(u+v)G_{j-1}^{(2)}]\lambda^{m-j},$$
$$ V_{21}^{(m)}= \sum_{j=0}^m [(G_{j-1,x}^{(2)}+
G_{j-1,x}^{(1)})-2(u-v)G_{j-1}^{(2)}]\lambda^{m-j}.$$

Then the compatibility condition of (2.4) and (2.5) is $U_{t_{m}}-
V_{x}^{(m)}+ [U,V^{(m)}]=0,$ which is equivalent to the hierarchy of
nonlinear evolution equations
$$u_{t_{m}} =  G_{m-1,xx}^{(2)} - 2(uG_{m-1}^{(2)})_{x}= 2G_{mx}^{(1)},$$
$$v_{t_{m}} = -G_{m-1,xx}^{(1)} - 2uG_{m-1,x}^{(1)} - 2(vG_{m-1}^{(2)})_{x} - 2vG_{m-1,x}^{(2)},$$
in brief,
$$(u_{t_{m}},v_{t_{m}})^{T}= X_{m},\,\,\ m \geq 0, \eqno(2.6)$$ and
$X_{j}=KG_{j-1}=JG_{j}.$ The first two nontrivial equations are

$$
\begin{array}{ll}
u_{t}  =  -2u_{x},\\
v_{t}  =  -2v_{x},
\end{array}\eqno(2.7)
$$
and
$$
\begin{array}{ll}
u_{t}  =  v_{xx}-2(uv)_{x},\\
v_{t}  =  u_{xx}+2uu_{x}-6vv_{x},
\end{array}\eqno(2.8)
$$
the second system (2.8) is our called the coupled Burgers equation.

Assume that (2.4) and (2.5) have two basic solutions $\psi =
(\psi_{1}, \psi_{2})^{T}$ and $\phi = (\phi_{1}, \phi_{2})^{T}.$ We
define a matrix W by
$$W= \frac{1}{2}(\phi \psi^{T} + \psi \phi^{T})\sigma = \begin{array}{ccc}
\left(
\begin{array}{cc}G & F \\
 H & -G \
\end{array}\right)
\end{array}, \,\,\ \sigma = \begin{array}{ccc}
\left(
\begin{array}{cc} 0 & -1 \\
 1 & 0 \
\end{array}\right)
\end{array}.\eqno(2.9)$$
By (2.4) and (2.5), we can find that
$$W_{x} = [U,W],\,\,\ W_{t_{m}}=[V^{(m)},W].\eqno(2.10)$$
Which implies that $\partial_{x} \ detW = 0, \,\,\
\partial_{t_{m}} \ detW = 0$, and (2.10) can be written as
$$
\begin{array}{ll}
G_{x} = (u+v)H - (u-v)F,\\
F_{x} = 2F(-\lambda +v) - 2(u+v)G,\\
H_{x}= 2(u-v)G -2(-\lambda + v)H,
\end{array}\eqno(2.11)
$$
and
$$\begin{array}{ll}
G_{t_{m}}& = HV_{12}^{(m)} - FV_{21}^{(m)},\\
F_{t_{m}}& = 2FV_{11}^{(m)} - 2GV_{12}^{(m)},\\
H_{t_{m}}&= 2GV_{21}^{(m)} -2HV_{11}^{(m)}.
\end{array}\eqno(2.12)$$
Select a,b as the form
$$a= \sum_{j=0}^N a_{j-1}\lambda^{N-j},\,\,\ b=\sum_{j=0}^N b_{j-1}\lambda^{N-j},$$
and suppose the functions $G, F, H$ have the following finite-order
polynomials in $\lambda $
$$\begin{array}{ll}
G = a_{x} + 2vb -2\lambda b ,\\
F = a_{x} - b_{x} + 2(u+v)b ,\\
H = -a_{x} - b_{x} +2(u-v)b.
\end{array}\eqno(2.13)$$
Substituting (2.13) into (2.11), we have
$$K\overline{G}=\lambda J\overline{G},\,\,\ \overline{G}=(a,b)^{T},$$
which is equivalent to
$$K\overline{G}_{j-1}= J\overline{G}_{j},\,\,\ J \overline{G}_{-1}=0,K\overline{G}_{N-1}=0,\,\,\
\overline{G}_{j}=(a_{j},b_{j})^{T}.\eqno(2.14)$$ It is easy to see
that the equation $J \overline{G}_{-1}=0$ has the general solution
$\overline{G}_{-1}= \alpha_{0}G_{-1} + \beta_{0}\widehat{G}_{-1},
\widehat{G}_{-1}= (1,0)^{T} \in KerJ.$ Where $\alpha_{0}$ and
$\beta_{0}$ are constants of integration, and we can obtain from
(2.1) and (2.14) that
 $$\overline{G}_{k}= \sum_{j=0}^{k+1}\alpha_{j}G_{k-j}+
\beta_{k+1}\widehat{G}_{-1}, \,\,\ -1 \leq k \leq N-1, \eqno(2.15)$$
where $\alpha_{0},\alpha_{1},\ldots ,\alpha_{k+1}$ and
$\beta_{k+1}$are constants of integration. Substituting (2.15) into
(2.14). yields the stationary
equation$$\sum_{j=0}^N\alpha_{0}X_{N-j}= \alpha_{0}X_{N}+
\alpha_{1}X_{N-1}+\ldots +\alpha_{N}X_{0}=0,\eqno(2.16)$$ this means
that expression (2.11) are existent. Without loss of generality, Let
$\alpha_{0}=1, \beta_{0}=0,$ from (2.14) and (2.15), we have
$$\overline{G}_{-1}=\begin{array}{ccc}
\left(
\begin{array}{cc} 1 \\ \,\ 1 \
\end{array}\right)
\end{array}, \,\,\
\overline{G}_{0}=\begin{array}{ccc}
\left(
\begin{array}{cc} u+ \alpha_{1} + \beta_{1} \\
 v + \alpha_{1} \
\end{array}\right)
\end{array},$$
$$\overline{G}_{1}=\begin{array}{ccc}
\left(
\begin{array}{cc} \frac{1}{2}v_{x}- uv - \alpha_{1}u + \alpha_{2} + \beta_{2} \\
 -\frac{1}{2}u_{x}-\frac{1}{2}u^{2} + \frac{3}{2}v^{2}+ \alpha_{1}v+ \alpha_{2} \
\end{array}\right)
\end{array}.$$

\section{Evolution of elliptic variables}
Now we suppose that the functions $G, F, H$ are finite-order
polynomials in $\lambda$ from (2.13)
$$G = \sum_{j=0}^{N+1}g_{j}\lambda^{N+1-j}, \,\,\ F = \sum_{j=0}^N f_{j}\lambda^{N-j}, \,\,\
H= \sum_{j=0}^N h_{j}\lambda^{N-j}, \eqno(3.1)$$ with
$$\begin{array}{ll}
g_{0} = -2b_{-1} ,\\
g_{j} = a_{j-2} + 2vb_{j-2} - 2b_{j-1} , \,\,\,\,\,\ 1 \leq j \leq N , \\
g_{N+1} = a_{N-1, x}+2vb_{N-1},\\
f_{j} = a_{j-1, x} - b_{j-1, x} + 2(u+v)b_{j-1} , \,\,\,\,\,\  1 \leq j \leq N ,  \\
h_{j} = -a_{j-1, x} - b_{j-1, x} +2(u-v)b_{j-1}, \,\,\,\,\,\  1 \leq
j \leq N .
\end{array}\eqno(3.2)$$
Therefore it is easy to calculate the first few members
$$\begin{array}{ll}
g_{0} = -2 , \,\,\  g_{1} = -2\alpha_{1}, \,\,\ g_{2} = 2u_{x} +
u^{2} +v^{2}
-2\alpha_{2},\\
f_{0}= 2(u+v), \,\,\  f_{1}=u_{x}- v_{x} +2v(u+v)+2\alpha_{1}(u+v),\\
f_{2}= \frac{1}{2}u_{xx} +\frac{1}{2}v_{xx}
-2u_{x}v-uv_{x}- \alpha_{1}u_{x}- \alpha_{1}v_{x}-3vv_{x}\\
 \,\,\,\,\,\,\,\,\,\,\,\ -u^{3}+3uv^{2} + 2\alpha_{1}uv + 2\alpha_{2}u-u^{2}v+3v^{3}+2\alpha_{1}v^{2}+ 2\alpha_{2}v,\\
h_{0}=2(u-v),\,\,\ h_{1}=-u_{x}-v_{x}+2v(u-v)+2\alpha_{1}(u-v),\\
h_{2}=\frac{1}{2}u_{xx}-\frac{1}{2}v_{xx}+2u_{x}v+uv_{x}+\alpha_{1}u_{x}-\alpha_{1}v_{x}-3vv_{x}\\
 \,\,\,\,\,\,\,\,\,\,\,\ -u^{3}+3uv^{2}+2\alpha_{1}uv+2\alpha_{2}u+u^{2}v-3v^{3}-2\alpha_{1}v^{2}-2\alpha_{2}v.
\end{array}\eqno(3.3)$$

We can write $F$ and $H$ as polynomials of $\lambda$ to define the
elliptic coordinates ${u_{i}}$ and ${v_{i}}$
$$F= 2(u+v)\prod_{i=1}^N (\lambda - u_{i}), \,\,\,\,\,\ H= 2(u-v)\prod_{i=1}^N (\lambda - v_{i}). \eqno(3.4)$$
By comparing the coefficients of $\lambda^{n-1}$ and
$\lambda^{n-2}$, we get
$$f_{1}=-2(u+v)\sum_{i=1}^N u_{i}, \,\,\,\ h_{1}=-2(u-v)\sum_{i=1}^N v_{i}, \eqno(3.5)$$
$$f_{2}= 2(u+v)\sum_{i < j} u_{i}u_{j}, \,\,\,\ h_{2} = 2(u-v)\sum_{i < j} v_{i}v_{j} .\eqno(3.6)$$
Thus from (3.3) and (3.5), we have
$$\begin{array}{ll}
-\frac{1}{2}\frac{(u-v)_{x}}{u+v}-v-\alpha_{1}= \sum\limits_{j=1}^N u_{j},\vspace{0.5cm}\\
\frac{1}{2}\frac{(u+v)_{x}}{u-v}-v-\alpha_{1}= \sum\limits_{j=1}^N
v_{j},$$
\end{array}\eqno(3.7)$$
from (3.3) and (3.6), we have
$$\begin{array}{ll}
\frac{1}{4}\frac{(u+v)_{xx}}{u+v}-\frac{\alpha_{1}}{2}\frac{(u+v)_{x}}{u+v}+\frac{3v^{2}}{2}+
\alpha_{1}v +
\alpha_{2}-\frac{u^{2}}{2}-v\frac{(u+v)_{x}}{u+v}-\frac{v_{x}}{2}=
\sum\limits_{i < j} u_{i}u_{j},\vspace{0.5cm}\\
\frac{1}{4}\frac{(u-v)_{xx}}{u-v}+\frac{\alpha_{1}}{2}\frac{(u-v)_{x}}{u-v}+\frac{3v^{2}}{2}+
\alpha_{1}v +
\alpha_{2}-\frac{u^{2}}{2}-v\frac{(u-v)_{x}}{u-v}-\frac{v_{x}}{2}=
\sum\limits_{i < j} v_{i}v_{j}.$$
\end{array}\eqno(3.8)$$
Consider the function $\det W$ which is a $(2N+2)$ th-order
polynomial in $\lambda$ with constant coefficients of the $x$- flow
and $t_{m}$-flow
$$-\det W=G^{2}+ FH= 4\prod_{j=1}^{2N+2}(\lambda-\lambda_{j})=R(\lambda).\eqno(3.9)$$
Substituting (3.1) into (3.9) and comparing the coefficients of
$\lambda^{2N+1}$ and $\lambda^{2N}$ yields
$$\begin{array}{ll}
2g_{0}g_{1}=-4\sum\limits_{j=1}^{2N+2} \lambda_{j},\\
2g_{0}g_{2}+ g_{1}^{2} + f_{0}h_{0}=4\sum\limits_{i <
j}\lambda_{i}\lambda_{j},
\end{array}\eqno(3.10)$$
together with (3.3) gives$$\alpha_{1}=-\frac{1}{2}\sum_{j=1}^{2N+2}
\lambda_{j} , \,\,\ \alpha_{2}=-\frac{1}{8}(\sum_{j=1}^{2N+2}
\lambda_{j})^{2}+\frac{1}{2}\sum_{i < j} \lambda_{i}\lambda_{j}.
\eqno(3.11)$$ From (3.9), we get that
$$G|_{\lambda=u_{k}}= \sqrt{R(u_{k})}, \,\,\ G|_{\lambda=v_{k}}= \sqrt{R(v_{k})}.\eqno(3.12)$$
Noticing (2.11) and (3.4), we obtain
$$F_{x}|_{\lambda = u_{k}}= -2(u+v)u_{k,x}\prod_{j=1,j\neq k}^N (u_{k}-u_{j})=-2(u+v)G|_{\lambda=u_{k}} ,\eqno(3.13)$$
$$H_{x}|_{\lambda = v_{k}}= -2(u-v)v_{k,x}\prod_{j=1,j\neq k}^N (v_{k}-v_{j})=-2(u-v)G|_{\lambda=v_{k}} .\eqno(3.14)$$
Hence we have the evolution of the elliptic coordinates along the
$x$ flows
$$u_{k,x}=\frac{\sqrt{R(u_{k})}}{\prod\limits_{j=1,j\neq
k}^N (u_{k}-u_{j})},\,\,\,\ 1 \leq k \leq N, \eqno(3.15)$$
$$v_{k,x}=- \frac{\sqrt{R(v_{k})}}{\prod\limits_{j=1,j\neq k}^N
(v_{k}-v_{j})},\,\,\,\ 1 \leq k \leq N. \eqno(3.16)$$ In a way
similar to the above expression , by using (2.12) and (3.4) we
arrive at
$$u_{k,t_{m}}=\frac{\sqrt{R(u_{k})}(u+v)^{-1}V_{12}^{(m)}|_{\lambda=u_{k}}}{\prod\limits_{i=1,i\neq k}^N (u_{k}-u_{i})} ,\,\,\,\ 1 \leq k \leq N, \eqno(3.17)$$
$$v_{k,t_{m}}=- \frac{\sqrt{R(v_{k})}(u-v)^{-1}V_{21}^{(m)}|_{\lambda=v_{k}}}{\prod\limits_{i=1,i\neq k}^N (v_{k}-v_{i})},\,\,\,\ 1 \leq k \leq N, \eqno(3.18)$$
and from (2.1), (2.2) and (2.5), we have
$$V_{12}^{(1)}|_{\lambda=u_{k}}= -2(u+v)u_{k}+(-u_{x}+v_{x})-2(u+v)v, \eqno(3.19)$$
$$V_{21}^{(1)}|_{\lambda=v_{k}}= -2(u-v)v_{k}+(v_{x}+u_{x})-2(u-v)v, \eqno(3.20)$$
$$
\begin{array}{lll}
V_{12}^{(2)}|_{\lambda=u_{k}}&=&-2(u+v)u_{k}^{2}+(u_{x}+v_{x}-2uv-2v^{2})u_{k}-\frac{1}{2}u_{xx}+3vv_{x} \\
&&-\frac{1}{2}v_{xx}+2u_{x}v+uv_{x}+u^{3}-3uv^{2}+u^{2}v-3v^{3},
\end{array}\eqno(3.21)$$
$$
\begin{array}{lll}
V_{21}^{(2)}|_{\lambda =v_{k}}&=& -2(u-v)v_{k}^{2}+(v_{x}-u_{x}-2uv+2v^{2})v_{k}-\frac{1}{2}u_{xx}+3vv_{x} \\
&&+\frac{1}{2}v_{xx}-2u_{x}v-uv_{x}+u^{3}-3uv^{2}-u^{2}v+3v^{3}.
\end{array}\eqno(3.22)$$

\section{Algebro-geometric solutions}
In the following, we will give the algebro-geometric solution of the
coupled Burgers  Equation(1.1). Returning to (3.9), we naturally
introduce the hyperelliptic curve $\mathcal{K}_{N}$ of arithmetic
genus $N$ defined by
$$\mathcal{K}_{N}: \,\,\,\ y^2 - R(\lambda) = 0. \eqno(4.1)$$ The curve
$\mathcal{K}_{N}$ can be compactified by joining two points at
infinity $P_{\infty_{\pm}}, \,\ P_{\infty_{+}} =
(P_{\infty_{-}})^{\ast}$. Still denoting its projective closure by
$\mathcal{K}_{N}$. Here we assume that the zeros $\lambda_{j}(j=1,
\ldots, 2N+2)$ of $R(\lambda)$ in (3.9) are mutually distinct, then
the hyperelliptic curve $\mathcal{K}_{N}$ becomes nonsingular.
According to the definition of $\mathcal{K}_{N}$, we can lift the
roots $\{u_{j}\}_{j=1, \ldots, N}, \,\,\ \{v_{j}\}_{j=1, \ldots, N}$
to $\mathcal{K}_{N}$ by introducing
$$\widehat{u}_{j}(x, t_{m})=(u_{j}(x, t_{m}),
G(u_{j}(x, t_{m}))), \eqno(4.2)$$
$$\widehat{v}_{j}(x, t_{m})=(v_{j}(x, t_{m}),
-G(v_{j}(x, t_{m}))), \eqno(4.3)$$ where $j=1, \ldots, N, \,\
(x,t_{m})\in \mathbb{R}^2$. Moreover, from (4.1) we know that $y^2 =
G^2 + FH$, that is $(y+G)(y-G)=FH$, then we can define the
meromorphic function $\phi(\cdot, x, t_{m})$ on $\mathcal{K}_{N}$
$$\phi(\cdot, x, t_{m})=\frac{y + G}{F} = \frac{H}{y - G}, \eqno(4.4)$$
where $P = (\lambda, y) \in \mathcal{K}_{N} \setminus
 \{P_{\infty_{\pm}}\}$. Hence the divisor of
 $\phi(P, x, t_{m})$ is
 $$(\phi(P, x, t_{m}))= \mathcal{D}_{\hat{\underline{v}}(x, t_{m})P_{\infty_{+}}}(P) -
 \mathcal{D}_{\hat{\underline{u}}(x, t_{m})P_{\infty_{-}}}(P), \eqno(4.5)$$
where
$$\hat{\underline{u}}(x, t_{m})= \{\hat{u}_{1}(x, t_{m}),
\ldots , \hat{u}_{N}(x, t_{m})\} \in Sym^{N}(\mathcal{K}_{N}),$$
$$
\hat{\underline{v}}(x, t_{m})= \{\hat{v}_{1}(x, t_{m}), \ldots ,
\hat{v}_{N}(x, t_{m})\} \in Sym^{N}(\mathcal{K}_{N}).$$ And the
branch of $y(\cdot)$ near $P_{\infty_{\pm}}$ is fixed according to
$$\lim\limits_{|z|\rightarrow \infty \atop P\rightarrow P_{\infty_{\pm}}}\frac{y(P)}{G(z)}= \mp 1.$$

Based on the definition of meromorphic function $\phi(\cdot, x,
t_{m})$ in (4.4), the spectral problem (2.4) and the auxiliary
problem (2.5), we can define the Baker-Akhiezer vector $\Psi(\cdot,
x, x_{0}, t_{m}, t_{m,0})$ on $\mathcal{K}_{N} \backslash
\{P_{\infty_{+}}, P_{\infty_{-}}\}$ by
$$\Psi(\cdot, x, x_{0}, t_{m}, t_{m,0}) = \begin{array}{ccc} \left(
\begin{array}{ccc}
 \psi_{1}(\cdot, x, x_{0}, t_{m}, t_{m,0})\\
\psi_{2}(\cdot, x, x_{0}, t_{m}, t_{m,0}) \
\end{array}\right)
\end{array}, \eqno(4.6)$$
where
$$\begin{array}{ll} \psi_{1}(P, x, x_{0}, t_{m}, t_{m,0}) =
&\exp(\int_{x_{0}}^{x}((-\lambda + v(x^{'},t_{m}))-(u(x^{'},t_{m})+
v(x^{'},t_{m})))\phi(P,x^{'},t_{m}))dx^{'}
\\ & + \int_{t_{m,0}}^{t_{m}}(V_{11}^{(m)}(\lambda, x_{0}, s) -
V_{12}^{(m)}(\lambda, x_{0}, s)\phi(P, x_{0}, s))ds), \end{array}
\eqno(4.7)$$
$$\psi_{2}(P,x,x_{0},t_{m},t_{m,0}) = -\psi_{1}(P,x,x_{0},t_{m},t_{m,0})\phi(P,x,t_{m}), \eqno(4.8)$$
with $P \in \mathcal{K}_{N}, (x,t_{m}), (x_{0}, t_{m,0}) \in
\mathbb{R}^2$.

In the following, we introduce the Riemann surface $\Gamma$ of the
hyperelliptic curve $\mathcal{K}_{N}$ and equip $\Gamma$ with a
canonical basis of cycles: $a_{1}, a_{2},\ldots, a_{N}; b_{1},
b_{2}, \ldots, b_{N}$ which are independent and have intersection
numbers as follows
$$a_{i}\circ a_{j}=0, \,\,\
b_{i}\circ b_{j}=0, \,\,\ a_{i}\circ b_{j}=\delta_{ij},
i,j=1,2,\ldots,N.$$

We will choose the following set as our basis
$$\tilde{\omega}_{l}=\frac{\lambda^{l-1}d\lambda}{\sqrt{R(\lambda)}},
\,\,\ l=1,2,\ldots,N,$$ which are linearly independent homomorphic
differentials from each other on $\Gamma$, and let
$$A_{ij}=\int_{a_{j}}\tilde{\omega}_{i}, \,\,\
B_{ij}=\int_{b_{j}}\tilde{\omega}_{i}.$$ It is possible to show that
the matrices $A=(A_{ij})$ and $B=(B_{ij})$ are $N \times N$
invertible period matrices $\cite{Siegel-1971,Griffiths-1994}$.
 Now we define the
matrices $C$ and $\tau$ by $C=(C_{ij})=A^{-1}, \,\,\
\tau=(\tau_{ij})=A^{-1}B$. Then the matrix $\tau$ can be shown to
symmetric $(\tau_{ij}=\tau_{ji})$ and it has a positive-definite
imaginary part (Im $\tau > 0$). If we normalize $\tilde{\omega}_{j}$
into the new basis $\omega_{j}$
$$\omega_{j}= \sum_{l=1}^N C_{jl}\tilde{\omega}_{l}, \,\,\,\,\ l=1,2, \ldots , N,$$
then we have
$$\int_{a_{j}}\omega_{j} = \sum_{l=1}^N C_{jl}\int_{a_{j}}\tilde{\omega}_{l} = \sum_{l=1}^N C_{jl}A_{li} = \delta_{ji},$$
$$\int_{b_{j}}\omega_{i} = \sum_{l=1}^N C_{jl}\int_{b_{j}}\tilde{\omega}_{l} = \sum_{l=1}^N C_{jl}B_{li} = \tau_{ji}.$$
Now we define the Abel-Jacobi coordinates
$$\rho_{j}^{(1)}(x, t_{m}) = \sum_{k=1}^N \int_{P_{0}}^{\hat{u}_{k}(x, t_{m})}\omega_{j}=
\sum_{k=1}^N\sum_{l=1}^N\int_{\lambda(P_{0})}^{u_{k}}C_{jl}
\frac{\lambda^{l-1}d\lambda}{\sqrt{R(\lambda)}}, \eqno(4.9)$$
$$\rho_{j}^{(2)}(x,t) = \sum_{k=1}^N \int_{P_{0}}^{\hat{v}_{k}(x, t_{m})}\omega_{j}=
\sum_{k=1}^N\sum_{l=1}^N\int_{\lambda(P_{0})}^{v_{k}}C_{jl}
\frac{\lambda^{l-1}d \lambda}{\sqrt{R(\lambda)}}, \eqno(4.10)$$
where $\lambda(P_{0})$ is the local coordinate of $P_{0}$. From
(3.15) and (4.9), we get
$$\partial_{x}\rho_{j}^{(1)}=
\sum_{k=1}^N \sum_{l=1}^N
C_{jl}\frac{u_{k}^{l-1}u_{kx}}{\sqrt{R(u_{k})}}=
\sum_{k=1}^N\sum_{l=1}^N
\frac{2C_{jl}u_{k}^{l-1}}{\prod\limits_{j=1,j\neq
k}^N(u_{k}-u_{j})},$$ which implies
$$\partial_{x}\rho_{j}^{(1)}= 2C_{jN}=\Omega_{j}^{(1)}, \,\,\,\
j=1,2,\ldots, N, \eqno(4.11)$$ With the help of the following
equality
$$
\sum_{k=1}^N\frac{u_{k}^{l-1}}{\prod\limits_{i=1,i \neq
k}^N(u_{k}-u_{i})}=\left \{
\begin{array}{ll}
\delta_{lN} ,& \textrm{$l= 1,2,\ldots,N$},
\\ \sum \limits_{i_{1}+i_{2}+\ldots
+i_{N}=l-N}u_{1}^{i_{1}}u_{2}^{i_{2}} \cdots u_{N}^{i_{N}},&
\textrm{$ l > N $} .
\end{array}
\right.\eqno(4.12)$$ In a similar way, we obtain from (4.9), (4.10),
(3.7), (3.15), (3.16), (3.17), (3.18), (3.19), (3.20):
$$\partial_{t}\rho_{j}^{(1)} = -2C_{j,N-1}+2\alpha_{1}C_{j,N} =
 \Omega_{j}^{(2)}, \,\,\,\ j=1,2,\ldots ,N, \eqno(4.13)$$
$$\partial_{x}\rho_{j}^{(2)}=-\Omega_{j}^{(1)}, \,\,\,\ j=1,2,\ldots ,N,\eqno(4.14)$$
$$\partial_{t}\rho_{j}^{(2)}=-\Omega_{j}^{(2)}, \,\,\,\ j=1,2,\ldots ,N.\eqno(4.15)$$

Let $\mathcal{T}$ be the lattice generated by $2n$ vectors
$\delta_{j}, \tau_{j}$, where $\delta_{j} = (\underbrace{0, \ldots ,
0}\limits_{j-1}, 1, \underbrace{0, \ldots, 0}\limits_{n-j})$ and
$\tau_{j}= \tau \delta_{j}$, the Jacobian variety of $\Gamma$ is
$\mathcal{J}= \mathbb{C}^n / \mathcal{T}$. On the basis of these
results, we obtain the following
$$\rho_{j}^{(1)}(x,t)=\Omega_{j}^{(1)}x + \Omega_{j}^{(2)}t + \gamma_{j}^{(1)}, \eqno(4.16)$$
$$\rho_{j}^{(2)}(x,t)=-\Omega_{j}^{(1)}x - \Omega_{j}^{(2)}t + \gamma_{j}^{(2)}, \eqno(4.17)$$
where $\gamma_{j}^{(i)}(i=1,2)$ are constants, and
$$\underline{\rho}^{(1)} = (\rho_{1}^{((1)}, \rho_{2}^{(1)} \ldots , \rho_{N}^{(1)})^{T}, \,\,\,\
\underline{\rho}^{(2)} = (\rho_{1}^{((2)}, \rho_{2}^{(2)}, \ldots ,
\rho_{N}^{(2)})^{T},$$
$$\underline{\Omega}^{(m)}=(\Omega_{1}^{(m)}, \Omega_{2}^{(m)}, \ldots ,
\Omega_{N}^{(m)})^{T}, \,\,\
\underline{\gamma}^{(m)}=(\gamma_{1}^{(m)}, \gamma_{2}^{(m)}, \ldots
, \gamma_{N}^{(m)})^{T}, \,\,\,\ m=1,2.$$ Now we introduce the Abel
map $\mathcal{A}(P): Div(\Gamma)\rightarrow \mathcal{J}$
$$\mathcal {A}(P)=\int_{P_{0}}^P \underline{\omega} , \,\,\,\ \underline{\omega}=(\omega_{1}, \omega_{2}, \ldots, \omega_{N})^T,$$
$$\mathcal {A}(\sum_{k}n_{k}P_{k})=\sum n_{k}\mathcal {A}(P_{k}), \,\,\ P, \,\ P_{k} \in \mathcal{K}_{N},$$
the Riemann theta function is defined as
$\cite{Gesztesy-2003,Siegel-1971,Griffiths-1994}$
$$\theta(P, D_{\hat{\underline{u}}(x, t_{m})}) = \theta(\underline{\Lambda} - \mathcal{A}(P) + \underline{\rho}^{(1)}),$$
$$\theta(P, D_{\hat{\underline{v}}(x, t_{m})}) = \theta(\underline{\Lambda} - \mathcal{A}(P) + \underline{\rho}^{(2)}),$$
where $\underline{\Lambda} = (\Lambda_{1},\ldots , \Lambda_{N})$ is
defined by
$$\Lambda_{j}=\frac{1}{2}(1 + \tau_{jj}) -
\sum\limits_{i=1,i\neq j}^N \int_{a_{i}} \omega_{i} \int_{Q_{0}}^P
\omega_{j}, \,\,\,\ j=1, \ldots , N.$$

In order to derive the algebro-geometric solution of coupled Burgers
equation(1.1), now we turn to the asymptotic properties of the
meromorphic function $\phi$ and Baker-Akhiezer function $\psi_{1}$.

{\bf Lemma 4.1.} Suppose that $u(x, t_{m}), v(x, t_{m}) \in
C^{\infty}(\mathbb{R}^2)$ satisfy the coupled Burgers equation
(1.1). Moreover, let $P=(\lambda, y) \in \mathcal{K}_{N}\backslash
\{P_{\infty_{\pm}}\}, \,\ (x, x_{0}) \in \mathbb{R}^2$. Then
$$\begin{array}{ll}
\phi(P) \underset{\zeta \rightarrow 0}{=} \left \{
\begin{array}{ll}
\frac{u-v}{2}\zeta + [\frac{(u-v)_{x}}{4} + \frac{v(u-v)}{2}]\zeta^2
+ O(\zeta^{3}) \,\,\ as \,\,\ P \rightarrow P_{\infty_{+}},
\\  \frac{-2}{u+v}\zeta^{-1} + \frac{2v}{u + v} - \frac{(u + v)_{x}}{(u + v)^2} + O(\zeta) \,\,\ as \,\,\ P
\rightarrow P_{\infty_{-}},
\end{array}
\right.
\end{array}
\eqno(4.18)$$ and
$$\begin{array}{ll}
\psi_{1}(P, x, x_{0}, t_{m}, t_{m,0}) \underset{\zeta \rightarrow
0}{=} \left \{
\begin{array}{ll}
\exp(-\zeta^{-1}(x-x_{0}) - 2\zeta^{-m-1}(t_{m} - t_{m,0}) + O(1))
\,\,\ as \,\,\ P \rightarrow P_{\infty_{+}},
\\ \exp(\zeta^{-1}(x-x_{0}) + 2\zeta^{-m-1}(t_{m} - t_{m,0})
+ O(1)) \,\,\ as \,\,\ P \rightarrow P_{\infty_{-}}.
\end{array}
\right.
\end{array}
\eqno(4.19)$$

{\bf Proof.} We first prove $\phi$ satisfies the Riccati-type
equations
$$\phi_{x}(P) + 2(-\lambda + v)\phi(P)- (u+v)\phi^2(P) + (u-v)
=0. \eqno(4.20)$$ The local coordinates $\zeta = \lambda^{-1}$ near
$P_{\infty_{\pm}}$ , from (4.4), (2.11), we have
$$\begin{array}{ll}
\phi_{x} &= \frac{G_{x}F - (y + G)F_{x}} {F^2} \\&=
\frac{(u+v)H-(u-v)F}{F} - \phi \frac{2F(-\lambda + v) - 2(u+v)G}{F}
\\&=-(u-v)-2(-\lambda +v)\phi + \frac{(u+v)H + 2(u+v)G\phi}{F},
\end{array}
\eqno(4.21)
$$
$$\phi^2 = \frac{y^2 + 2yG + G^2}{F^2}
=\frac{2G^2 + FH + 2yG}{F^2} = \frac{2G\phi + H}{F}, \eqno(4.22)$$
according to (4.21) and (4.22), we have (4.20). And then, inserting
the ansatz $\phi \underset{\lambda \rightarrow 0}{=}
\phi_{1}\lambda^{-1} + \phi_{2}\lambda^{-2} + O(\lambda^{-3})$ into
(4.20), we get the first line of (4.18). Inserting he ansatz $\phi
\underset{\lambda \rightarrow 0}{=} \phi_{-1} \lambda + \phi_{0} +
\phi_{1}\lambda^{-1} + O(\lambda^{-2})$ into (4.20), we get the
second line of (4.18). In the following, we will prove (4.19). From
(4.7) and (4.18)
$$\begin{array}{ll}
& \,\,\,\,\,\  \exp(\int_{x_{0}}^x ((-\lambda - v(x^{'}, t_{m})) -
(u(x^{'}, t_{m}) + v(x^{'}, t_{m}))\phi)dx^{'}) \\&=
\exp(\int_{x_{0}}^x ((-\zeta^{-1} - v) - (u+v)\phi)dx^{'})
\\&\underset{\zeta \rightarrow 0}{=} \left \{
\begin{array}{ll}
\exp(\int_{x_{0}}^x (-\zeta^{-1} - v) -(u+v)(\frac{u-v}{2}\zeta +
[\frac{(u-v)_{x}}{4} + \frac{v(u-v)}{2}]\zeta^2 + O(\zeta^{3}))
\,\,\ as \,\,\ P \rightarrow P_{\infty_{+}},
\\ \exp(\int_{x_{0}}^x (-\zeta^{-1} - v) -
(u+v)(\frac{-2}{u+v}\zeta^{-1} + \frac{2v}{u + v} - \frac{(u +
v)_{x}}{(u + v)^2} + O(\zeta)) \,\,\ as \,\,\ P \rightarrow
P_{\infty_{-}},
\end{array}
\right.\\& \underset{\zeta \rightarrow 0}{=}\left \{
\begin{array}{ll}
\exp(-\zeta^{-1}(x - x_{0}) + O(1)) \,\,\ as \,\,\ P \rightarrow
P_{\infty_{+}},
\\ (\frac{u(x) + v(x)}{u(x_{0}) + v_{x_{0}}} + O(\zeta))exp(\zeta^{-1}(x - x_{0}) + O(1) \,\,\ as \,\,\ P \rightarrow
P_{\infty_{-}}.
\end{array}
\right.
\end{array}
\eqno(4.23)$$
From (4.1) and (3.9), we have
$$\begin{array}{ll}
y &= \mp\sqrt{R(\lambda)}\\& =\mp 2 \prod\limits_{j=1}^{2N +
2}(\lambda - \lambda_{j})^{\frac{1}{2}}\\& = \mp 2\zeta^{-N-1}
\prod\limits_{j=1}^{2N + 2}(1 - \lambda_{j}\zeta)^{\frac{1}{2}}
\\& \underset{\zeta \rightarrow 0}{=} \mp
2\zeta^{-N-1}\prod\limits_{j=1}^{2N+2} (1 + \epsilon_{1}\zeta +
\epsilon_{2}\zeta^2 + O(\zeta^3)) \,\,\,\ as \,\,\ P\rightarrow
P_{\infty_{\pm}},
\end{array}
\eqno(4.24)$$ where
$\epsilon_{1}=-\frac{1}{2}\sum\limits_{j=1}^{2N+2}\lambda_{j}, \,\,\
\epsilon_{2} = \frac{1}{2}\sum\limits_{j < k} \lambda_{j}\lambda_{k}
- \frac{1}{8}(\sum\limits_{j=1}^{2N+2} \lambda_{j})^2$. From (3.1)
and (3.4), we can derive
$$
\begin{array}{ll}
F^{-1}&= \frac{1}{2(u+v)}\prod\limits_{j=1}^N \frac{1}{\lambda -
u_{j}} \\&= \frac{1}{2(u + v)}\zeta^N \prod\limits_{j=1}^N
\frac{1}{1 - u_{j}\zeta}\\& \underset{\zeta \rightarrow
0}{=}\frac{1}{2(u + v)}\zeta^N(1 + \sum\limits_{j=1}^N u_{j}\zeta +
O(\zeta^2)) \,\,\ as \,\ P \rightarrow P_{\infty_{\pm}},
\end{array} \eqno(4.25)$$
combining (4.18), (4.26), (4.27) we have
$$\begin{array}{ll}
& \,\,\,\,\ \exp(\int_{t_{m,0}}^{t_{m}}(V_{11}^{(m)}(\lambda, x_{0},
s) - V_{12}^{(m)}(\lambda, x_{0}, s)\phi(P, x_{0}, s))ds) \\& =
\exp(\int_{t_{m,0}}^{t_{m}} (V_{11}^{(m)} - V_{12}^{(m)}\frac{y +
G}{F})ds)
\\& = \exp(\int_{t_{m,0}}^{t_{m}}(\frac{-y}{F}V_{12}^{(m)} +
\frac{F_{t_{m}}}{2F})ds) \\&\underset{\zeta \rightarrow 0}{=} \exp
(\int_{t_{m,0}}^{t_{m}}(\pm\zeta^{-N-1}(1+
O(\zeta))\frac{\sum\limits_{j=0}^m [(G_{j-1, x}^{(2)} -
G_{j-1,x}^{(1)}) - 2(u +
v)G_{j-1}^{(2)}]\lambda^{m-j}}{\sum\limits_{j=0}^N
f_{j}\lambda^{N-j}} + \frac{u_{t_{m}}+ v_{t_{m}}}{2(u+ v)}ds)
\\& \underset{\zeta \rightarrow 0}{=}
\exp(\int_{t_{m,0}}^{t_{m}}(\pm\zeta^{-N-1}(1+ O(\zeta))
\frac{\sum\limits_{j=0}^m [(G_{j-1, x}^{(2)} - G_{j-1,x}^{(1)}) -
2(u + v)G_{j-1}^{(2)}]\zeta^{j-m}}{\sum\limits_{j=0}^N
f_{j}\zeta^{j-N}})+ \frac{u_{t_{m}}+ v_{t_{m}}}{2(u+ v)}ds)\\&
\underset{\zeta \rightarrow 0}{=} \exp
(\int_{t_{m,0}}^{t_{m}}(\pm\zeta^{-m-1}(1+
O(\zeta))\frac{\sum\limits_{j=0}^m [(G_{j-1, x}^{(2)} -
G_{j-1,x}^{(1)}) - 2(u +
v)G_{j-1}^{(2)}]\zeta^{j}}{\sum\limits_{j=0}^N f_{j}\zeta^{j}} +
\frac{u_{t_{m}}+ v_{t_{m}}}{2(u+ v)}ds)\\& \underset{\zeta
\rightarrow 0}{=} \exp (\int_{t_{m,0}}^{t_{m}}(\pm\zeta^{-m-1}(1+
O(\zeta))\frac{(G_{-1, x}^{(2)} - G_{-1,x}^{(1)}) - 2(u +
v)G_{-1}^{(2)}}{f_{0}} + \frac{u_{t_{m}}+ v_{t_{m}}}{2(u+ v)}ds)\\&
\underset{\zeta \rightarrow 0} {=}\left \{
\begin{array}{ll}
\exp(-\zeta^{-m-1}(t_{m} - t_{m,0}) + O(1)) \,\,\ as \,\,\ P
\rightarrow P_{\infty_{+}},
\\ \exp(\zeta^{-m-1}(t_{m} - t_{m,0}) + O(1)) \,\,\ as \,\,\ P \rightarrow
P_{\infty_{-}},
\end{array}
\right.
\end{array} \eqno(4.26)$$
according to the definition of $\psi_{1}$ in (4.7), (4.25) and
(4.26), we can obtain (4.19). $\Box$

Next, we shall derive the representation of $\phi, \psi_{1},
\psi_{2}, u(x,t_{m}), v(x, t_{m})$ in term of the Riemann theta
function. Let $\omega_{P_{\infty_{+},P_{\infty_{-}}}}^{(3)}$ be the
normalized differential of the third kind holomorphic on
$\mathcal{K}_{N} \backslash \{P_{\infty_{+}}, P_{\infty_{-}}\}$ with
simples at $P_{\infty_{+}}$ and $P_{\infty_{-}}$ and residues $1$
and $-1$ respectively,
$$\omega_{P_{\infty_{+},P_{\infty_{-}}}}^{(3)} =
\frac{1}{y}\prod\limits_{j=1}^N (\lambda - \lambda_{j})d\lambda
\underset{\zeta \rightarrow 0}{=}(\pm\zeta^{-1}+O(1))d\zeta \,\,\ as
P \rightarrow P_{\infty_{\pm}}, \eqno(4.27)$$ here the constants
$\{\lambda_{j}|\lambda_{j}\in\mathbb{C}, j=1,\ldots , N\}$ are
uniquely determined by the normalization
$$\int_{a_{j}}\omega_{P_{\infty_{+},P_{\infty_{-}}}}^{(3)}= 0, \,\,\ j=1,\ldots, N, \eqno(4.28)$$
and $\zeta$ in (4.30) denotes the local coordinate $\zeta =
\lambda^{-1}$ for $P$ near $P_{\infty_{\pm}}$. Moreover,
$$\int_{Q_{0}}^P \omega_{P_{\infty_{+},P_{\infty_{-}}}}^{(3)} \underset{\zeta \rightarrow 0}{=}
\ln(\zeta)-\ln \omega_{0} + O(\zeta) \,\,\ as P \rightarrow
P_{\infty_{+}}, \eqno(4.29)$$ and $$\int_{Q_{0}}^P
\omega_{P_{\infty_{+},P_{\infty_{-}}}}^{(3)} \underset{\zeta
\rightarrow 0}{=} -(\ln(\zeta)-\ln \omega_{0} + O(\zeta)) \,\,\ as P
\rightarrow P_{\infty_{-}}. \eqno(4.30)$$

Let$\omega_{P_{\infty_{\pm}},r}^{(2)}, \,\ r \ln N_{0}$, be
normalized differentials of the second kind with a unique pole at
$P_{\infty_{\pm}}$, and principal part is $\zeta^{-2-r}d\zeta$ near
$P_{\infty_{\pm}}$, satisfying
$$\int_{a_{j}} \omega_{P_{\infty_{\pm}},r}^{(2)} = 0, j=1,\ldots,N,$$
then we can define $\Omega_{0}^{(2)}$ and $\Omega_{m-1}^{(2)}$ by
$$\Omega_{0}^{(2)} = \omega_{P_{\infty_{-}},0}^{(2)} - \omega_{P_{\infty_{+}},0}^{(2)},   \eqno(4.31)$$
$$\Omega_{m-1}^{(2)} =
\sum\limits_{l=0}^{m}\alpha_{m}(l+1)(\omega_{P_{\infty_{-}},l}^{(2)}-
\omega_{P_{\infty_{+}},l}^{(2)}), \eqno(4.32)$$ where
$\alpha_{m-1-l}, \,\ j=0, \ldots, m-1$ are the integral constants in
(2.15), so we have
$$\int_{a_{j}}\Omega_{0}^{(2)}=0, \int_{a_{j}}\Omega_{m-1}^{(2)}, \,\ j=1, \ldots, N, \eqno(4.33)$$
$$\int_{Q_{0}}^P \Omega_{0}^{(2)} \underset{\zeta \rightarrow 0}{=}
\mp(\zeta^{-1} + e_{0,0} + O(\zeta)) \,\ as P\rightarrow
P_{\infty_{\pm}}, \eqno(4.34)$$
$$\int_{Q_{0}}^P \Omega_{m-1}^{(2)}\underset{\zeta \rightarrow 0}{=}
\mp(\sum\limits_{l=0}^{m}\alpha_{m-1}\zeta^{-1-l} + e_{m,0} +
O(\zeta)) \,\ as P\rightarrow P_{\infty_{\pm}}, \eqno(4.35)$$ for
some constants $e_{0,0}, e_{m, 0} \in \mathbb{C}$.

If $D_{\underline{\widehat{u}}(x, t_{m})}$ or
$D_{\underline{\widehat{v}}(x, t_{m})}$ in (4.5) is assumed to be
nonspecial$\cite{Gesztesy-2003}$, then according to Riemann's
theorem$\cite{Gesztesy-2003,Griffiths-1994}$, the definition and
asymptotic properties of the meromorphic function $\phi(P, x,
t_{m})$, $\phi(P, x, t_{m})$ has expressions of the following type
$$\phi(P, x, t_{m}) = N(x,t_{m})\frac{\theta(P, D_{\underline{\widehat{v}}(x, t_{m})})}
{\theta(P, D_{\underline{\widehat{u}}(x,
t_{m})})}\exp(\int_{Q_{0}}^P
\omega_{P_{\infty_{+},P_{\infty_{-}}}}^{(3)}), \eqno(4.36)$$ where
$N(x, t_{m})$ is independent of $P \in \mathcal{K}_{N}$.

{\bf Theorem 4.1.} Let $P=(\lambda, y) \in \mathcal{K}_{N}\backslash
{P_{\infty_{\pm}}}, \,\ (x, t_{m}), (x_{0}, t_{m,0}) \in \Omega$,
where $\Omega \subseteq \mathbb{R}^2$ is open and connected. Suppose
$u(\cdot, t_{m}), v(\cdot, t_{m})\in C^{\infty}(\Omega), u(x,
\cdot), v(x, \cdot)\in C^{1}(\Omega), \,\ x\in \mathbb{R}, t_{m} \in
\mathbb{R}$, satisfy the equation (1.1), and assume that
$\lambda_{j}, 1\leq j \leq 2N+2$ in (3.9) satisfy $\lambda_{j} \in
\mathbb{C}$ and $\lambda_{j} \neq \lambda_{k}$ for $j \neq k$.
Moreover, suppose that $D_{\underline{\widehat{u}}}$ or
equivalently, $D_{\underline{\widehat{v}}}$, is nonspecial for $(x,
t_{m}) \in \Omega$. Then
$$\phi(P, x, t_{m}) = N(x,t_{m})\frac{\theta(P, D_{\underline{\widehat{v}}(x, t_{m})})}
{\theta(P, D_{\underline{\widehat{u}}(x,
t_{m})})}\exp(\int_{Q_{0}}^P
\omega_{P_{\infty_{+},P_{\infty_{-}}}}^{(3)}),  \eqno(4.37)$$
$$
\begin{array}{ll}
&\psi_{1}(P, x, x_{0}, t_{m}, t_{m,0}) =
\frac{\theta(P_{\infty_{+}}, D_{\underline{\widehat{u}}(x_{0},
t_{m,0})})\theta(P, D_{\underline{\widehat{u}}(x,
t_{m})})}{\theta(P_{\infty_{+}}, D_{\underline{\widehat{u}}(x,
t_{m})})\theta(P, D_{\underline{\widehat{u}}(x_{0}, t_{m,0})})}
\\& \,\,\,\,\,\,\,\,\,\,\,
\times \exp((\int_{Q_{0}}^P \Omega_{0}^{(2)} +
e_{0,0})(x-x_{0})+(\int_{Q_{0}}^P \Omega_{m}^{(2)} +
e_{m,0})(t_{m}-t_{m,0})),
\end{array} \eqno(4.38)$$
$$\begin{array}{ll}
&\psi_{2}(P, x, x_{0}, t_{m}, t_{m,0}) = N(x,t_{m}) \frac{\theta(P,
D_{\underline{\widehat{v}}(x, t_{m})})\theta(P_{\infty_{+}},
D_{\underline{\widehat{u}}(x_{0}, t_{m,0})})\theta(P,
D_{\underline{\widehat{u}}(x, t_{m})})}{\theta(P,
D_{\underline{\widehat{u}}(x, t_{m})})\theta(P_{\infty_{+}},
D_{\underline{\widehat{u}}(x, t_{m})})\theta(P,
D_{\underline{\widehat{u}}(x_{0}, t_{m,0})})}
\\& \,\,\,\,\,\,\,\,\,\,\,  \times
\exp((\int_{Q_{0}}^P \Omega_{0}^{(2)} +
e_{0,0})(x-x_{0})+(\int_{Q_{0}}^P \Omega_{m}^{(2)} +
e_{m,0})(t_{m}-t_{m,0}))
\\& \,\,\,\,\,\,\,\,\,\,\, \times \exp(\int_{Q_{0}}^P
\omega_{P_{\infty_{+}},P_{\infty_{-}}}^{(3)}).
\end{array} \eqno(4.39)$$
Finally, $u(x, t_{m})$ is of the form
$$u(x, t_{m}) = \omega_{0}N(x,t_{m})\frac{\theta(P_{\infty_{+}},
D_{\underline{\widehat{v}}(x, t_{m})}}{\theta(P_{\infty_{+}},
D_{\underline{\widehat{u}}(x, t_{m})})} -
\frac{1}{\omega_{0}N(x,t_{m})}\frac{\theta(P_{\infty_{-}},
D_{\underline{\widehat{u}}(x, t_{m})}}{\theta(P_{\infty_{-}},
D_{\underline{\widehat{v}}(x, t_{m})})}, \eqno(4.40)$$ and $v(x,
t_{m})$ is of the form
$$v(x, t_{m}) = -
\frac{1}{\omega_{0}N(x,t_{m})}\frac{\theta(P_{\infty_{-}},
D_{\underline{\widehat{u}}(x, t_{m})}}{\theta(P_{\infty_{-}},
D_{\underline{\widehat{v}}(x, t_{m})})}
-\omega_{0}N(x,t_{m})\frac{\theta(P_{\infty_{+}},
D_{\underline{\widehat{v}}(x, t_{m})}}{\theta(P_{\infty_{+}},
D_{\underline{\widehat{u}}(x, t_{m})})},  \eqno(4.41)$$ and $N(x,
t_{m})$ is determined by
$$
\begin{array}{lll} \omega_{0}N(x,t_{m})\frac{\theta(P_{\infty_{+}},
D_{\underline{\widehat{v}}(x, t_{m})}}{\theta(P_{\infty_{+}},
D_{\underline{\widehat{u}}(x, t_{m})})} +
\frac{1}{\omega_{0}N(x,t_{m})}\frac{\theta(P_{\infty_{-}},
D_{\underline{\widehat{u}}(x, t_{m})}}{\theta(P_{\infty_{-}},
D_{\underline{\widehat{v}}(x, t_{m})})} \\ = \partial_{x}\ln
\frac{\theta(P_{\infty_{+}}, D_{\underline{\widehat{u}}(x_{0},
t_{m,0})})\theta(P_{\infty_{-}}, D_{\underline{\widehat{u}}(x,
t_{m})})}{\theta(P_{\infty_{-}}, D_{\underline{\widehat{u}}(x_{0},
t_{m,0})})\theta(P_{\infty_{+}}, D_{\underline{\widehat{u}}(x,
t_{m})})} - 2e_{0,0}. \end{array} \eqno(4.42)
$$
{\bf proof.} We start with the proof of the theta function
representation (4.38).Without loss of generality, it suffices to
treat the special case of (2.15) when $\alpha_{0}=2, \alpha_{k}=0, 1
\leq k \leq N$. First, we assume
$$u_{j}(x, t_{m}) \neq u_{k}(x, t_{m}), \,\ for j\neq k,
and \,\ (x, t_{m}) \in \widetilde{\Omega}  \eqno(4.43)$$ for
appropriate $\widetilde{\Omega} \subseteq \Omega$, and define the
right-hand side of (4.38) to be $\widetilde{\psi}_{1}$. In order to
prove $\psi = \widetilde {\psi}$, we investigate the local zeros and
poles of $\psi_{1}$. From (3.4), (3.15), (3.16), (3.17), (3.18),
(4.4), we have
$$\begin{array}{ll}
(u(x^{'}, t_{m})+ v(x^{'}, t_{m}))\phi(P, x^{'}, t_{m}) &\underset{P
\rightarrow \widehat{u}_{j}(x^{'}, t_{m})}{=}
\frac{y(\widehat{u}_{j}(x^{'}, t_{m}))}{\prod\limits_{k=1, k \neq
j}(u_{j}(x^{'}, t_{m}) - u_{j}(x^{'}, t_{m}))}\frac{1}{\lambda -
u_{j}(x^{'}, t_{m})}\\& \underset{P \rightarrow
\widehat{u}_{j}(x^{'}, t_{m})}{=} \frac{u_{j,x^{'}}}{y-u_{j}(x^{'},
t_{m})}\\&\underset{P \rightarrow \widehat{u}_{j}(x^{'},
t_{m})}{=}-\partial_{x^{'}}\ln(y-u_{j}(x^{'}, t_{m})) + O(1).
\end{array}
\eqno(4.44)$$ And similarly
$$
V_{12}^{(m)}(\lambda, x_{0}, s)\phi(P, x_{0}, s) \underset{P
\rightarrow \widehat{u}_{j}(x_{0}, s)}{=} -
\partial_{s} \ln(y - u_{j}(x_{0},s)) + O(1), \eqno(4.45)$$
then (4.44) and (4.45) together with (4.7) yields
$$
\psi_{1}(P, x, x_{0}, t_{m}, t_{m,0}) = \left \{
\begin{array}{ll}
(\lambda - u_{j}(x,t_{m}))O(1), \,\,\ as \,\,\ P \rightarrow
\widehat{u}_{j}(x, t_{m}) \neq \widehat{u}_{j}(x_{0}, t_{m,0})
 \\
O(1), \,\,\ as \,\,\,\ P \rightarrow \widehat{u}_{j}(x, t_{m}) =
\widehat{u}_{j}(x_{0}, t_{m,0})
\\ (\lambda - u_{j}(x_{0},t_{m,0}))^{-1}O(1), \,\,\ as
\,\,\ P \rightarrow \widehat{u}_{j}(x_{0}, t_{m,0}) \neq
\widehat{u}_{j}(x, t_{m})
\end{array}
\right. \eqno(4.46)$$ where $P=(\lambda, t_{m}) \in \mathcal{K}_{N},
(x,t_{m}), (x_{0}, t_{m,0}) \in \widetilde{\Omega}$ and $O(1) \neq
0$. Hence $\psi_{1}$ and $\widetilde{\psi}_{1}$ have identical zeros
and poles on $\mathcal{K}_{N} \backslash \{P_{\infty_{\pm}}\}$,
which are all simple by hypothesis (4.43). It remains to study the
behavior of $\psi_{1}$ and $\widetilde{\psi}_{1}$ near
$P_{\infty_{\pm}}$, by (4.19), (4.34), (4.35), (4.38), we can easy
find that $\psi_{1}$ and $\widetilde{\psi}_{1}$ share the same
singularities and zeros, and the Riemann-Roch-type
uniqueness$\cite{Gesztesy-2003}$ proves that $\psi_{1} =
\widetilde{\psi}_{1}$, hence (4.38) holds subject to (4.43).

Substituting (4.29), (4.30) into (4.36) and comparing with (4.18),
we obtain
$$u(x, t_{m})- v(x,t_{m}) = 2\omega_{0}N(x,t_{m})
\frac{\theta(P_{\infty_{+}}, D_{\underline{\widehat{v}}(x,
t_{m})})}{\theta(P_{\infty_{+}}, D_{\underline{\widehat{u}}(x,
t_{m})})}, \eqno(4.47)
$$
$$u(x, t_{m})+ v(x,t_{m}) = \frac{-2}{\omega_{0}N(x,t_{m})}
\frac{\theta(P_{\infty_{-}}, D_{\underline{\widehat{u}}(x,
t_{m})})}{\theta(P_{\infty_{-}}, D_{\underline{\widehat{v}}(x,
t_{m})})}, \eqno(4.48)
$$
according to (4.48), we have (4.40) and (4.41), and $\psi_{2}$ in
(4.39) from $\psi_{2}=-\phi\psi_{1}$. Reexamining the asymptotic
behavior of $\psi_{1}$ near $P_{\infty_{-}}$ yields
$$\begin{array}{ll} \psi_{1}(P, x, x_{0}, t_{m}, t_{m,0}) =
& \exp(\int_{x_{0}}^{x}(-v(x^{'},t_{m})dx^{'} + O(\zeta))\\& \times
\exp(\zeta^{-1}(x-x_{0})+\zeta^{-m-1}(t_{m}-t_{m,0})+O(1)).
\end{array}
\eqno(4.49)$$ On the other hand, according to (4.38), (4.34),
(4.35), we have
$$
\begin{array}{ll}
&\psi_{1}(P, x, x_{0}, t_{m}, t_{m,0}) =
\frac{\theta(P_{\infty_{+}}, D_{\underline{\widehat{u}}(x_{0},
t_{m,0})})\theta(P_{\infty_{-}}, D_{\underline{\widehat{u}}(x,
t_{m})})}{\theta(P_{\infty_{+}}, D_{\underline{\widehat{u}}(x,
t_{m})})\theta(P_{\infty_{-}}, D_{\underline{\widehat{u}}(x_{0},
t_{m,0})})}
\\& \,\,\,\,\,\,\,\,\,\,\,
\times \exp((\zeta^{-1} + 2e_{0,0} + O(\zeta))(x-x_{0}) +
(\zeta^{-m-1} + 2e_{m,0} + O(\zeta))(t_{m}-t_{m,0})),
\end{array} \eqno(4.50)$$
a comparison of (4.49) and (4.50) proves (4.42).

Hence, we prove this theorem on $\widetilde{\Omega}$. The extension
of all these results from $\widetilde{\Omega}$ to $\Omega$ follows
by continuity of the Abel map and the nonspecial nature of
$D_{\underline{\widehat{u}}}$ or $D_{\underline{\widehat{v}}}$ on
$\Omega$. $\Box$

Therefore, the algebro-geometric solution of (1.1) is (4.40) and
(4.41) for $m=1$.

\section{Acknowledgements}

The Project is in part supported by the Natural Science Foundation
of China (Grant No. 11271008), the First-class Discipline of
University in Shanghai and the Shanghai Univ. Leading Academic
Discipline Project (A.13-0101-12-004).


\end{document}